\documentclass[letterpaper, 10 pt, conference]{ieeeconf}  % Comment this line out if you need a4paper

\IEEEoverridecommandlockouts                              % This command is only needed if 
\usepackage{times} % assumes new font selection scheme installed
\usepackage{cite}
\usepackage{amsmath,amssymb,amsfonts}
\usepackage{algorithmic}

\usepackage{graphicx}
\graphicspath{{./images/}}
\DeclareGraphicsExtensions{.pdf,.png,.jpg}

\usepackage{empheq}
\usepackage{hyperref}
\hypersetup{colorlinks,citecolor=blue,linkcolor=[rgb]{0,0,0.4},urlcolor=[rgb]{0,0,0.4},pdftitle={CDC paper Marcantony 2023}}

\usepackage{textcomp}
\def\BibTeX{{\rm B\kern-.05em{\sc i\kern-.025em b}\kern-.08em
    T\kern-.1667em\lower.7ex\hbox{E}\kern-.125emX}}
\usepackage{mathtools}
\usepackage{bbm}
\usepackage{thmtools}
\usepackage{flushend}
\usepackage{tikz}
\usetikzlibrary{positioning,quotes,angles} % , quotes,calc,patterns
\usepackage{color}
\usepackage{dsfont}

\newcommand{\rline}{{\mathbb R}}

\newcommand{\bbm}[1]{\left[\begin{matrix} #1 \end{matrix}\right]}
\newcommand{\sbm}[1]{\left[\begin{smallmatrix} #1 \end{smallmatrix}\right]}

\newcommand{\bbmi}[2]{\begin{bmatrix*}[#1] #2 \end{bmatrix*}}

\newcommand{\rfb}[1]{\mbox{\rm
   (\ref{#1})}\ifx\undefined\stillediting\else:\fbox{$#1$}\fi}

\renewenvironment{proof}{\vspace{.1cm}\noindent{\sc
    Proof.}\hspace{0.10cm}\,\,}{$\hfill\Box$\vspace{.3cm}} 

\newtheorem{theorem}{Theorem}[section] 
 
\newtheorem{assumption}{Assumption}[section] 
 
\newtheorem{lemma}[theorem]{Lemma} 
\newtheorem{proposition}[theorem]{Proposition}

\newcommand{\sign}{\text{ sign}}
\newcommand{\degree}{ ^{\circ}}

\title{\LARGE \bf Range-Only Bearing Estimator for Localization and Mapping} %Relative Bearing Estimation of Static Landmark via Range-only Measurements
\author{Matteo Marcantoni$^{1}$, Bayu Jayawardhana$^{2}$, and Kerstin Bunte$^{1}$% <-this % stops a space
\thanks{This work was supported in part by the Project SMART-AGENTS of the Research Programme Smart Industry which is (partly)
financed by the Dutch Research Council (NWO) under Project 18024.
}% <-this % stops a space
\thanks{$^{1}$Matteo Marcantoni and Kerstin Bunte are with the Bernoulli Institute, Faculty of Science and Engineering, University of Groningen, 9747AG Groningen, The Netherlands (email: m.marcantoni@rug.nl; kerstin.bunte@gmail.com).} %and Kerstin Bunte are ; kerstin.bunte@gmail.com
\thanks{$^{2}$Bayu Jayawardhana is with the Engineering and Technology Institute Groningen, Faculty of Science and Engineering, University of Groningen, 9747AG Groningen, The Netherlands (email: b.jayawardhana@rug.nl).}
}

\begin{document}

\maketitle
\thispagestyle{empty}
\pagestyle{empty}
\begin{abstract}
Navigation and exploration within unknown environments are typical examples in which simultaneous localization and mapping (SLAM) algorithms are applied. 
When mobile agents deploy only range sensors without bearing information, the agents must estimate the bearing using the online distance measurement for the localization and mapping purposes. 
In this paper, we propose a scalable dynamic bearing estimator to obtain the relative bearing of the static landmarks in the local coordinate frame of a moving agent in real-time. 
Using contraction theory, we provide convergence analysis of the proposed range-only bearing estimator and present upper and lower-bound for the estimator gain. 
Numerical simulations demonstrate the effectiveness of the proposed method.   
\end{abstract}

%\begin{IEEEkeywords}
%keywords here
%\end{IEEEkeywords}
\section{Introduction}
\label{sec:introduction}
Navigation and exploration in unknown GPS-denied environments (e.g. underwater, indoor, etc.) are typical applications of the simultaneous localization and mapping (SLAM) algorithms, which can provide crucial information for the positioning and control of mobile agents \cite{ukf2020,Guo2020,ocean2006}. 
%In this context, range-only simultaneous localization and mapping (RO-SLAM) pertains to the problem of estimating the position of a moving agent and of various landmarks in the local or global coordinate frame based on the use of range sensor systems, which give only the relative distance information.
In this context, range-only simultaneous localization and mapping (RO-SLAM) pertains to the problem of estimating the position of a moving agent and of various landmarks in the global or local coordinate frame based on the use of range sensor systems, which only provide relative distance information.
This is the case when ranging sensors, such as acoustic or radio devices, are employed to sense the environment. 
%Examples of popular range-only sensor systems are ultra-wideband (UWB) devices as used in the relative localization of unmanned aerial vehicles \cite{Guo2017} and of automated guided vehicles \cite{Monica2021}.
Examples of popular range-only sensor systems are ultra-wideband (UWB) devices as used in the relative localization of unmanned aerial vehicles (UAVs) \cite{Guo2017} and of automated guided vehicles (AGVs) \cite{Monica2021}.
The main challenge in the RO-SLAM problem is the real-time estimation of the bearing information based only on the distance measurement signals.
%\mm{Certainly, the main challenge in RO-SLAM is to estimate (in real-time) relative bearing information with only distance measurement signals.}
The combination of bearing and range information can then provide the relative position of the landmarks with respect to the moving agent.  

Various methods have been proposed in the literature to solve the problem using extended Kalman filter SLAM (EKF-SLAM) \cite{ocean2006,ekf2023,ekf2015,ekf2013,ekf2009}, unscented Kalman filter SLAM (UKF-SLAM) \cite{ukf2020} and particle filter SLAM (PF-SLAM) \cite{pf2022}. 
%These methods work well locally, which can introduce a practical issue in the deployment of such methods, namely, they require a good initial estimate of the relative position in order to guarantee the convergence of the algorithms.
These methods work well locally, which can introduce a practical issue in their deployment. 
Namely, they require a good initial estimate of the relative position in order to guarantee the convergence of the algorithms.
In general, there are two approaches for the initialization in literature: the delayed and undelayed approaches \cite{ekf2023}.
In the first approach, the application of the estimation filter is delayed until consistent estimates of the positions for the landmarks are obtained through multiple measurements (such as, the triangulation techniques \cite{ekf2015,ocean2006}). 
This approach can lead to convergence problems particularly when the delay is set too large. 
The second approach relies on the availability and evaluation of multiple probable hypotheses in order to allow for immediate initialization of the landmarks' relative position and for subsequent application of the estimation filter. 
The optimization process during such pre-processing step and the recursive computation in these SLAM filters can lead to high computational load \cite{ekf2023,ekf2013,ekf2009,pf2022}.

%RO-SLAM can also be considered as a relative localization problem when the local coordinate frame of the moving agent is used instead of the global one.
RO-SLAM can also be considered as a relative localization problem when the local coordinate frame of the moving agent is used over the global one.
Swarming and formation control are typical applications that consider this setting \cite{Guo2017}.  
In these cases, estimating the relative displacement between moving agents and/or static landmarks is a pre-requisite to deploy formation or swarming algorithms. 
When beacons are deployed as landmarks, source localization algorithms can also be used to solve the estimation problem \cite{indiveri2012,Batista2011,Dandach2009}. 
Specifically, nonlinear least-squares Gaussian-Newton (NLS-GS) method \cite{Guo2017}, recursive least-squares filter \cite{indiveri2012}, Kalman filter \cite{Batista2011} and a continuous-time adaptive estimator \cite{Dandach2009} have been proposed in the literature.

In this paper, we propose a dynamic bearing estimator design method. 
The estimator design requires only one estimator state variable per landmark (so that it is scalable) and it guarantees semi-global exponential convergence. 
The design method does not require multi-hypothesis testing, delayed approaches or previous source localization techniques. 
The update of the estimator state is based on comparing the expected distance (obtained from the current agent's displacement and current bearing estimate) and the current distance measurement. 
The convergence of the estimator is analyzed via recent results of contraction theory \cite{contr2019,contr1998}. 
%We provide the lower and upper-bound of the estimator gain to guarantee the exponential convergence and depends on the initial condition of the estimator. 
We provide the lower and upper-bound of the estimator gain to guarantee the exponential convergence.%, which also depends on the initial condition of the estimator.}
%We show also the existence of another attractive point when the agent is moving in a straight line and the semi-global convergence property of the estimator when the agent does not move in a straight line.
We furthermore show the existence of two attractive points when the agent is moving in a straight line and the semi-global convergence property of the estimator when the agent moves in a curved trajectory. %\kb{winding} 
The efficacy of the proposed method is demonstrated via numerical simulations.

The paper is organized as follows. 
Preliminaries and problem formulation are presented in \autoref{sec:preliminaries}. 
In \autoref{sec:range-only} we describe the design of our observer, in \autoref{subsec:contraction_analysis} and \ref{subsec:convergence_analysis} we analyze its contraction and convergence properties respectively, and in \autoref{sec:numerical_sim} we present simulation results.

\section{Preliminaries and Problem Formulation}
\label{sec:preliminaries}
In this section, we first define the estimation problem at hand and briefly present existing results on contraction systems theory that are central to the analysis of our main results.
For the sake of clarity, we present the setup and problem formulation for a single landmark, while the extension to multiple landmarks will be discussed later towards the end of \autoref{subsec:convergence_analysis}.

\subsection{Systems description and problem formulation}
\label{subsec:problem_formulation}
For describing the dynamics of a moving agent and a static landmark, let us consider the following discrete-time system with a constant sampling time $T$:
\begin{empheq}[left = \empheqlbrace]{align}
    \label{eq:sys}
     \notag p(k+1) &= p(k) + T u(k) \enspace, \\
     y(k) &= \|l^* - p(k)\| \enspace,
\end{empheq}
%\begin{equation}
%    \label{eq:sys}
%    \left\{ 
%    \begin{array}{rl}
%        p(k+1) &= p(k) + T u(k) ,
%        \\
%    %    l(k+1) &= l^*,\\
%        y(k) &= \|l^* - p(k)\| \enspace,
%    \end{array}\right.
%\end{equation}
where $k$ denotes the time, $p(k)\in \rline^2$ the position of the agent, $u(k)\in \rline^2$ the velocity input, $y(k)\in \rline$ the output of the system (i.e. the relative distance measurement), and $l^*\in \rline^2$ is the true position of a static landmark, which is unknown to the agent.
The positions are defined in the local coordinate frame of the agent.
As mentioned in the introduction, our design goal is to design a dynamic estimator of the bearing $\theta(k)$ which can be combined with the distance information $y(k)$ to obtain the (relative) position $l^*-p(k)=y(k)\sbm{\cos(\theta^*(k))\\ \sin(\theta^*(k))}$, where $\theta^*(k)$ is the true relative bearing information.
From the RO-SLAM context, our design problem is closely related to the mapping task as we assume that we have good knowledge of the agent position $p(k)$ in its local coordinate frame at any given time $k$. 
In practice, this assumption can be met by taking the initial point of the agent as the origin, its initial orientation defines the orientation of the frame and its displacement can be measured by fusing information from the on-board odometer and IMU sensor systems. 

By definition, the true relative bearing $\theta^*(k)\in \rline$ between the agent and the static landmark can be rewritten as:
\begin{align}
    \label{eq:bearing_true} &\theta^*(k) = \tan^{-1}\left(\dfrac{[(l^* - p(k))]_{2}}{[(l^* - p(k))]_{1}}\right)\enspace,
\end{align}
where $[\cdot]_{1}$, $[\cdot]_{2}$ denote the first and second components of a 2D vector, respectively. 
In our range-only bearing estimation problem we define the bearing estimation error $e(k)$ as: 
\begin{align}
\label{eq:est_error} &e(k) = \theta^*(k) - \theta(k) \enspace,
\end{align}
and the problem can be defined formally as follows. 

%\vspace{0.1cm}
\noindent{\bf Range-Only Bearing Estimation Problem:} 
For the system \eqref{eq:sys}, design a dynamic estimator 
\begin{equation}
    \label{eq:estimator1}
    \theta(k+1) = f\left(\theta(k),p(k),y(k)\right) \enspace,
\end{equation}
with continuously differentiable $f$, such that the bearing estimation error $e(k)\to 0$ as $k\to\infty$. 
%\vspace{0.1cm}

In the above problem formulation, the dynamic estimator \eqref{eq:estimator1} relies only on the use of distance measurement $y(k)$ and current position information $p(k)$ to update the bearing estimator state. 
Before introducing our proposed estimator, let us we briefly present known results from contraction theory in the following subsection that we will use to analyze the convergence property. % of our dynamic estimator.

\subsection{Contraction theory}
\label{subsec:contraction_preliminaries}
Consider a discrete-time non linear system 
\begin{align}
    \label{eq:general_sys} 
    x(k+1) = f(x(k),k) \enspace,
\end{align}
where $x(k)\in\mathcal{X}\subseteq\rline^n$ is the state and $f(x(k),k)$ is a continuously differentiable function. 
In the seminal paper \cite{contr1998} the contraction theory (for continuous- and discrete-time systems) pertains to the study of the convergence property of any trajectories of \eqref{eq:general_sys} to each other. 
The system \eqref{eq:general_sys} is said to be {\em contracting} if for any two trajectories $x_1(k)$ and $x_2(k)$ of \eqref{eq:general_sys} starting from two different initial state $x_1(0)\neq x_2(0)\in\mathcal X$, there exists $0<\lambda<1$, such that
\[
\|x_1(k)-x_2(k)\| \leq \lambda^{k} \|x_1(0)-x_2(0)\|, \quad \forall k\geq 0 \enspace.
\]
%This property can be established by analyzing the variational system
This property is established via variational system analysis:
\begin{equation}
    \label{eq:diff_general_sys} \delta x(k+1) = %\left. 
    \underbrace{\dfrac{\partial f}{\partial x}(x(k),k)}_{=:A(x(k),k)}%\right |_{x(k)} 
    \delta x(k) \enspace,
\end{equation}
where $\delta x(k)$ denotes the variational state. 

%\vspace{0.1cm}
\begin{lemma}[Lohmiller and Slotine in \cite{contr1998}] 
\label{lemma:contraction}
If the variational system \eqref{eq:diff_general_sys} is uniformly stable in $\mathcal X$ (i.e. the eigenvalues of $A(x(k),k)$ lie uniformly inside the unit disc), then \eqref{eq:general_sys} is contracting. 
\end{lemma}
%\vspace{0.1cm}

We will use Lemma \ref{lemma:contraction} above to establish the convergence of the estimator state $\theta(k)$ to the true relative bearing $\theta^*(k)$. 
In this regard, we can express \eqref{eq:sys}-\eqref{eq:estimator1} into \eqref{eq:general_sys}, where the input velocity signal $u(k)$ is considered as an external signal. 
In particular, our estimator design is independent to the design of any controller to regulate the mobile agent motion in the plane. 
As will be shown later, the update of the estimator requires the mobile agent to move. % around. 
%This implies that if no motion control is being applied to the mobile agent for completing a certain task, we need to displace the mobile robot by applying an appropriate input signal to map the environment, as discussed at the end of next section. 
This implies that if no motion control is being applied to the mobile agent for completing a certain task, we need to displace the mobile robot by applying an appropriate input signal to map the environment.

\section{Relative bearing estimator design}
\label{sec:range-only}
Without loss of generality, we assume that the local coordinate frame is aligned with the global one. 
Let us denote $v(k)$ as a unit vector that gives the direction from the current agent position to the estimated landmark position, and $w(k)$ as a unit vector orthogonal to $v(k)$, i.e.
\begin{equation}
    \label{eq:unit_vectors}
    v(k) = \bbm{\cos(\theta(k))\\ \sin(\theta(k))}, \ \text{ and }
    w(k) = \bbmi{r}{\sin(\theta(k))\\ -\cos(\theta(k))} 
    %\bbm{\hspace*{0.2cm}\sin(\theta(k))\\ -\cos(\theta(k))}
    \enspace.
\end{equation}
Furthermore, we assume that the input vector $u(k)$ is not collinear to $v(k)$, i.e. the inner product $\left\langle u(k), w(k) \right\rangle \neq 0$, for all $k \geq 0$ and the following assumption on the sampling time $T$ and on the input signal $u(k)$ hold. 

%\vspace{0.1cm}
\begin{assumption}
    \label{ass:small_ang}
    The mobile agent displacement as in \eqref{eq:sys} is sufficiently small, such that the small-angle approximation holds, i.e. 
    \begin{align*}
        \theta(k+1)&-\theta(k) \approx \sin(\theta(k+1) -\theta(k)) \\
        &= \dfrac{T}{y(k)} \left\langle u(k), w(k) \right\rangle, \quad \forall k\geq 0 .
    \end{align*}
\end{assumption}
%\vspace{0.1cm}

This assumption can be satisfied for sufficiently small $T \|u(k)\|$, i.e. we can use a combination of high sampling rate and small magnitude of the control input $u(k)$. 

Based on the estimation problem introduced in \autoref{sec:preliminaries} our proposed relative bearing estimator design is given as: 
%\begin{equation}
%    \label{eq:estimator}     
%    \left\{ 
%    \begin{array}{rl}
%        \theta(k+1) &= \theta(k) + T\langle u(k),w(k)\rangle \\
%        & \hspace{1.1cm} + \gamma \sign\left(\langle u(k),w(k)\rangle \right)  \beta(k), \\
%        l(k) & = y(k)v(k) + p(k) \enspace ,
%    \end{array}\right.
%\end{equation}
\begin{empheq}[left = \empheqlbrace]{align}
\label{eq:estimator}
    \notag \theta(k+1)   =& \theta(k) + \dfrac{T}{y(k)}\langle u(k),w(k)\rangle \\
        &\hspace{0.65cm} +\gamma \sign\left(\langle u(k),w(k)\rangle \right)  \beta(k) \enspace, \\
    \notag l(k)               =& y(k)v(k) + p(k) \enspace,
\end{empheq}
where $\gamma$ is the gain of the observer, $\langle \cdot,\cdot \rangle$ is the inner product operator, $\sign(\cdot)$ denotes the sign function, and $\beta(k)$ is the correction term given by:
\begin{align}
    \nonumber & \beta(k) = y^2(k+1) - \|l(k) - p(k+1) \|^2 \\
    \nonumber & =  \|l^* - p(k+1) \|^2  - \|l(k) - p(k+1) \|^2 \\
    \label{eq:correction-term} & = \|l^* - p(k) - T u(k) \|^2  - \|l(k) - p(k) - T u(k) \|^2.
\end{align}

Hence, $\beta(k)$ gives the mismatch between the expected distance given the current bearing estimate and the actual one when the agent has moved to the next position. 
Since $l(k) - p(k) = y(k)v(k)$ as in \eqref{eq:estimator}, \eqref{eq:correction-term} can be expressed as: 
\begin{align}
    \label{eq:correction-term1} \beta(k) =  \|l^* - p(k) - T u(k)\|^2 - \left\|y(k)v(k) - T u(k)\right\|^2.
\end{align}

\subsection{Contraction analysis}
\label{subsec:contraction_analysis}
Before delving into the analysis, let us review the overall system comprising of the plant and the estimator as follows:
\begin{empheq}[left = \empheqlbrace]{align} 
    \label{eq:closed-loop} 
    %\left\{ 
    %\begin{array}{rl}
    \notag  p(k+1) &= p(k) + T u(k), \\ %    l(k+1) &= l^*\\
    \theta(k+1) &= \theta(k) + \dfrac{T}{\|l^* - p(k)\|}\left\langle u(k),w(k) \right\rangle \\
    \notag  &\hspace{1.15cm} + \gamma \sign\left(\langle u(k),w(k)\rangle \right)  \beta(k) \enspace, 
    %\end{array}\right.
\end{empheq} 
where $x(k) = \bbm{p(k) & \theta(k)}^{\top}$ is the state of the overall system and $\beta(k)$ is defined as in \eqref{eq:correction-term1}.
Firstly, we analyze the contraction property of the second equation of the overall system \eqref{eq:closed-loop} in this subsection.
Subsequently, we provide the analysis of the steady-state trajectory to which all trajectories converge.

%\vspace{0.1cm} 
\begin{proposition}
\label{prop:contraction}
Suppose that there exists $0<c<1$ such that the input $|\langle u(k),w(k)\rangle| \geq  c\|u(k)\|$ for all $k\geq 0$, i.e. the direction of $u(k)$ is within a cone with the axis of $w(k)$ for all time. 
If, for a given input signal $u(k)$, the estimator gain $\gamma > 0$ satisfies 
\begin{empheq}[left = \empheqlbrace]{align}
    \label{eq:gain_inequality}%    \left\{
    \begin{array}{rl}%\notag
    \gamma &> \dfrac{1}{2c\|l^* - p(k)\|^2} \enspace, \\
    %   \gamma &< \dfrac{2 - T\|u(k) \|}{2T\|l^* - p(k)\| \left|\langle u(k),w(k) \rangle\right|},
    \gamma &< \dfrac{2\|l^* - p(k)\| - T\|u(k) \|}{2T\|l^* - p(k)\|^2 \|u(k)\|} \quad \forall k \geq 0\enspace,
\end{array}%\right.
\end{empheq}
then the estimator dynamics in \eqref{eq:closed-loop} is contracting, i.e.
\begin{align}
\label{eq:estimator_prop}
    \nonumber \theta(k+1) = \theta(k) &+ \dfrac{T}{\|l^* - p(k)\|}\left\langle u(k),w(k) \right\rangle \\
    &+ \gamma \sign(\langle u(k),w(k)\rangle)  \beta(k) \enspace.
\end{align}
Moreover, if $T\|u\|_\infty<\frac{2c}{1+c}\min\left(\|l^*-p(k)\|\right)$ holds then there exists $\gamma >0$ satisfying \eqref{eq:gain_inequality}.
\end{proposition}
%\vspace{0.1cm}

\begin{proof}
Since the sign function is non-differentiable the proof concerns two cases.
We first analyze the case where $\sign(\left\langle u(k),w(k)\right\rangle) = 1$ and thus $\langle u(k),w(k) \rangle \geq c\|u(k)\|$.  
Note that the estimator in \eqref{eq:estimator_prop} can be expressed as a nonlinear system as in \eqref{eq:general_sys}.
In order to show that \eqref{eq:estimator_prop} is a contracting system, we can analyze the corresponding variational system following \autoref{lemma:contraction}.
Let us denote the right-hand side of \eqref{eq:estimator_prop} by $f(\theta(k),k)$.
By computing its Jacobian, we obtain that
\begin{align}
    \label{eq:derivative_theta} 
    %\left.\begin{array}{rl}
    \notag\dfrac{\partial f}{\partial \theta}(\theta(k),k) = 1 &+ \dfrac{T}{\|l^* - p(k)\|} \left\langle  u(k),v(k) \right\rangle \\ %&\hspace{0.6cm} 
        &- 2 \gamma  T \|l^* - p(k)\| \left\langle u(k),w(k) \right\rangle \enspace,
    %\end{array} \right.
\end{align}
where $v(k)$ and $w(k)$ are defined in \eqref{eq:unit_vectors}. 
Hence, in order to apply \autoref{lemma:contraction} we need to verify that %it is necessary that
\begin{equation}
    \label{eq:cond-contr}
    -1< \dfrac{\partial f}{\partial \theta}(\theta(k),k) <1
\end{equation}
holds for all $k\geq 0$. 
Substituting \eqref{eq:derivative_theta} into the above inequality we get arrive at the following inequalities: 
\begin{equation}
    \label{eq:contraction}  
    \left\{
    \begin{array}{l}
    \dfrac{T \left\langle u(k),v(k) \right\rangle}{\|l^* - p(k)\|}  - 2 \gamma T \|l^* - p(k)\| 
      \left\langle u(k),w(k) \right\rangle < 0 ,\\
    \dfrac{T \left\langle u(k),v(k) \right\rangle}{\|l^* - p(k)\|} - 2 \gamma T \|l^* - p(k)\| 
      \left\langle u(k),w(k) \right\rangle > -2 .
    \end{array}\right.
\end{equation}
\if{0}
\begin{empheq}[left = \empheqlbrace]{align}%{equation}
    \label{eq:contraction}  %\left\{
    \begin{array}{rl}
    \dfrac{T \left\langle u(k),v(k) \right\rangle}{\|l^* - p(k)\|}  - 2 \gamma T \|l^* - p(k)\| 
      \left\langle u(k),w(k) \right\rangle &< 0 ,\\ 
    \dfrac{T \left\langle u(k),v(k) \right\rangle}{\|l^* - p(k)\|} - 2 \gamma T \|l^* - p(k)\| 
      \left\langle u(k),w(k) \right\rangle &> -2 .
    \end{array}%\right.
\end{empheq}
\fi
By rearranging the inequalities above, it follows that  
\begin{empheq}[left = \empheqlbrace]{align}%{equation}
    \label{eq:contraction1} %\left\{ 
    %\begin{array}{rl}
    \notag\gamma &> \dfrac{ \left\langle u(k),v(k) \right\rangle}{2 \|l^* - p(k)\|^2 \left\langle u(k),w(k) \right\rangle} \enspace,\\
    \gamma &< \dfrac{2 \|l^* - p(k)\| + T \left\langle u(k),v(k)\right\rangle}{2 T \|l^* - p(k)\|^2 \left\langle u(k),w(k) \right\rangle} \enspace.
    %\end{array}%\right.
\end{empheq}%{equation}
%By noting that $-\|u(k)\| \leq \left\langle u(k),v(k) \right\rangle \leq \|u(k)\|$, we conclude that if \eqref{eq:gain_inequality} holds then it follows from the first inequality in \eqref{eq:gain_inequality} that 
%\begin{align*}
 %   \gamma &> \dfrac{ 1}{2c \|l^* - p(k)\|} > \dfrac{\|u(k)\|}{2\|l^* - p(k)\|\langle u(k),w(k)\rangle} \\
 %   & \geq \dfrac{\langle u(k),v(k)\rangle}{2T\|l^* - p(k)\|\langle u(k),w(k)\rangle}
 %   %\gamma &> \dfrac{ 1}{2c \|l^* - p(k)\|} \geq \dfrac{\|u(k)\|}{2\|l^* - p(k)\|\langle u(k),w(k)\rangle} \\
 %   %& \geq \dfrac{\langle u(k),v(k)\rangle}{2\|l^* - p(k)\|\langle u(k),w(k)\rangle},
%\end{align*}
Since $-\|u(k)\| \leq \left\langle u(k),v(k) \right\rangle \leq \|u(k)\|$ we conclude that if \eqref{eq:gain_inequality} holds it follows from the first inequality in \eqref{eq:gain_inequality} that 
\begin{align*}
    %\gamma &> \dfrac{ 1}{2c \|l^* - p(k)\|} > \dfrac{\|u(k)\|}{2\|l^* - p(k)\|\langle u(k),w(k)\rangle} \\
    %& \geq \dfrac{\langle u(k),v(k)\rangle}{2T\|l^* - p(k)\|\langle u(k),w(k)\rangle}
    \gamma > \dfrac{ 1}{2c \|l^* - p(k)\|^2} 
    &\geq \dfrac{\|u(k)\|}{2\|l^* - p(k)\|^2 \langle u(k),w(k)\rangle} \\
    &\geq \dfrac{\langle u(k),v(k)\rangle}{2\|l^* - p(k)\|^2 \langle u(k),w(k)\rangle}
\end{align*}
holds for all $k$ and from the second inequality in \eqref{eq:gain_inequality} that 
\begin{align*}
   \gamma %&
   < \dfrac{2\|l^* - p(k)\| - T\|u(k) \|}{2T\|l^* - p(k)\|^2 \|u(k)\|} 
   %\\&
   \leq \dfrac{2\|l^* - p(k)\| + T\langle u(k),v(k)\rangle}{2T\|l^* - p(k)\|^2 \langle u(k),w(k)\rangle}
\end{align*}
also holds for all $k$. 
Hence, the contraction condition in \eqref{eq:contraction1} (or in \eqref{eq:contraction}) holds when \eqref{eq:gain_inequality} is satisfied.
By \autoref{lemma:contraction} the estimator \eqref{eq:estimator_prop} is contracting.

Analogously, for the case of $\sign\left(\left\langle u(k),w(k)\right\rangle \right) = -1$, i.e. $\langle u(k),w(k) \rangle \leq  -c\|u(k)\|$, equation \eqref{eq:derivative_theta} becomes:
\begin{align}
    \label{eq:derivative_theta1} %\left.
    %\begin{array}{rl}
    \notag \dfrac{\partial f}{\partial \theta}(\theta(k),k) = 1 &+ \dfrac{T \left\langle  u(k),v(k) \right\rangle}{\|l^* - p(k)\|} \\
        %&\hspace{0.6cm} 
        &+ 2 \gamma  T \|l^* - p(k)\| \left\langle u(k),w(k) \right\rangle\enspace.
    %\end{array}\right.
\end{align}
Thus, the contraction condition \eqref{eq:cond-contr} is satisfied when 
%\if{0}
\begin{equation}
    \label{eq:contraction_neg}  
    \left\{
    \begin{array}{l}
    \dfrac{T \left\langle u(k),v(k) \right\rangle}{\|l^* - p(k)\|}  + 2 \gamma T \|l^* - p(k)\| \left\langle u(k),w(k) \right\rangle < 0,\\ 
    \dfrac{T \left\langle u(k),v(k) \right\rangle}{\|l^* - p(k)\|} + 2 \gamma T \|l^* - p(k)\| \left\langle u(k),w(k) \right\rangle > -2.
    \end{array} \right.
\end{equation}
%\fi
\if{0}
\begin{empheq}[left = \empheqlbrace]{align}%\begin{equation}
    \label{eq:contraction_neg} %\left\{%. 
    %\begin{array}{rl}
    \dfrac{T \left\langle u(k),v(k) \right\rangle}{\|l^* - p(k)\|}  + 2 \gamma T \|l^* - p(k)\| \left\langle u(k),w(k) \right\rangle &< 0 \enspace,\\ \notag
    \dfrac{T \left\langle u(k),v(k) \right\rangle}{\|l^* - p(k)\|} + 2 \gamma T \|l^* - p(k)\| \left\langle u(k),w(k) \right\rangle &> -2 \enspace.
    %\end{array}
\end{empheq}
\fi
By rewriting these inequalities we get:% have 
\begin{empheq}[left = \empheqlbrace]{align}%\begin{equation}
    \label{eq:contraction_neg2} %\left\{%. 
    %\begin{array}{rl}
    \notag \gamma &< - \dfrac{ \langle u(k),v(k) \rangle}{2 \|l^* - p(k)\|^2 \left\langle u(k),w(k) \right\rangle}\enspace,\\ 
    \gamma &> - \dfrac{2\|l^* - p(k)\| + T \langle u(k),v(k)\rangle}{2 T \|l^* - p(k)\|^2 \left\langle u(k),w(k) \right\rangle}\enspace.
    %\end{array} \right.
\end{empheq} %equation}
Since $\sign\left(\left\langle u(k),w(k)\right\rangle\right) = -1$ the inequalities become 
\begin{empheq}[left = \empheqlbrace]{align}%\begin{equation}
    \label{eq:contraction_bounds} %\left\{%. 
    %\begin{array}{rl}
    \notag \gamma &> \dfrac{ \langle u(k),v(k) \rangle}{2 \|l^* - p(k)\|^2 \left|\left\langle u(k),w(k) \right\rangle\right|}\enspace,\\
    \gamma &< \dfrac{2\|l^* - p(k)\| + T \langle u(k),v(k) \rangle}{2 T \|l^* - p(k)\|^2 \left|\left\langle u(k),w(k) \right\rangle\right|}\enspace.
%\end{array} %\right.
\end{empheq} %equation}
Following a similar argumentation as before, it can be shown that \eqref{eq:gain_inequality} implies \eqref{eq:contraction_bounds}. 

Finally, for the admissibility of $\gamma$ it is necessary that
\begin{equation*}
   \dfrac{ 1}{2c\|l^* - p(k)\|^2 } < \dfrac{2\|l^* - p(k)\| - T \|u(k)\|}{2 T \|l^* - p(k)\|^2\|u(k)\|} \enspace.
\end{equation*}
%This inequality is satisfied when
%\begin{equation}
 %  \label{eq:gain_existence}T \| u(k)\| < \frac{c}{2(1+c)}
%\end{equation}
%holds for all $k$, i.e $T\|u\|_\infty< \frac{c}{2(1+c)}$.
This inequality is satisfied when
\begin{equation}
\label{eq:gain_existence}T \| u(k)\| < \frac{2c}{1+c}\|l^* - p(k)\| \quad \text{holds } \forall k \enspace,
\end{equation}
%holds for all $k$, 
which is the case for $T\|u\|_\infty< \frac{2c}{1+c}\min\left(\|l^* - p(k)\|\right)$.
\end{proof}

As shown in \autoref{prop:contraction} the sampling time and the input signal $u$ should be taken sufficiently small for the admissibility of the estimator gain $\gamma$. 
This is in line with \autoref{ass:small_ang} where a sufficiently small $T\|u(k)\|$ is required. %assumed. 
Since the denominators in \eqref{eq:gain_inequality} depend on $\|l^*-p(k)\|^2$ the estimator gain becomes larger the closer the mobile agent gets to the landmark.
% the closer the mobile agent to the landmark the larger the estimator gain should be. 
However, this is compensated by the requirement that $T\|u(k)\|$ must be small in order to ensure the validity of small-angle approximation. 

%As a side remark, 
Note that for a specific motion of the agent where $\left\langle u(k),v(k) \right\rangle = 0\ \forall k \geq 0$, i.e. the agent is always moving perpendicularly to the estimated landmark position $l(k)$, one can exploit \eqref{eq:derivative_theta} to design an adaptive gain $\gamma(k)$ for the estimator, such that \eqref{eq:cond-contr} is always satisfied. 
That is: %For example
\begin{align*}
    %\left.\begin{array}{rl}
    \dfrac{\partial f}{\partial \theta}(\theta(k),k)  = 0 &\Rightarrow  2 \gamma(k) T \|l^* - p(k)\| \|u(k)\| = 1 \\
%\end{align*}
%\begin{align*}
    & \Rightarrow \gamma(k) = \dfrac{1}{2T\|l^* - p(k)\| \|u(k)\|} \enspace,
    %\end{array}\right.
\end{align*}
where $\gamma(k)$ is the estimator gain at %the 
time step $k$.

\subsection{Convergence analysis}
\label{subsec:convergence_analysis}
The next step in our analysis is to investigate whether the contractivity of the estimator in \eqref{eq:estimator_prop} implies that the estimated landmark position $l(k)$ converges to the true position $l^*$ as $k \to \infty$.
%In order tTo show this %result we present 
We demonstrate this separately for both, the case of straight and curved motion of the mobile agent, starting with the latter.
%Let us start with the latter as follows.

%\vspace{0.2cm}
\begin{proposition}
    \label{prop:convergence_no-straight}
    Assume that the hypotheses in \autoref{prop:contraction} hold for some $c,\gamma > 0$ so that \eqref{eq:estimator_prop} is contracting. 
    %If the agent is not moving on a straight trajectory (e.g. $u(k+1)\neq \epsilon(k) u(k)$ for any $\epsilon(k)\in \rline$ and for all $k\geq 0$), then the estimated landmark position $l(k)$ converges to the true landmark position $l^*$ as $k\to \infty$, i.e. $e(k)\to 0$.
    If the agent is not moving on a straight trajectory then the estimated landmark position $l(k)$ converges to the true landmark position $l^*$ as $k\to \infty$, i.e. $e(k)\to 0$.
\end{proposition}
%\vspace{0.2cm}

\begin{proof}
By the construction of the estimator one of the admissible trajectory of \eqref{eq:estimator_prop} is the case when $e=0$. 
In this case the corresponding steady-state trajectory $\theta_{\text{ss}}(k)$ satisfies:
\begin{equation}
    \label{eq:steady-state-theta}
    %\theta_{\text{ss}}(k+1) = \theta_{\text{ss}}(k) + \dfrac{T}{\|l^*-p(k)\|} \left\langle u(k), \bbmi{r}{\sin(\theta_{\text{ss}}(k))\\ -\cos(\theta_{\text{ss}}(k))} \right\rangle \enspace,
    \theta_{\text{ss}}(k+1) = \theta_{\text{ss}}(k) + \dfrac{T}{\|l^*-p(k)\|} \left\langle u(k), \sbm{\sin(\theta_{\text{ss}}(k))\\ -\cos(\theta_{\text{ss}}(k))} \right\rangle \enspace,
\end{equation}
which is invariant for all time steps, i.e. $l_{\text{ss}}(k)=l^*$ for all $k\geq 0$. 
By the contraction property of \eqref{eq:estimator_prop} all trajectories converge exponentially to each other; since $\theta(k) \to \theta_{\text{ss}}(k)$ as $k\to\infty$ when the initial error $e(0)\neq 0$. %exponentially 
%\mm{By the contraction property of \eqref{eq:estimator_prop}, all trajectories converge exponential to each other; hence $\theta(k) \to \theta_{\text{ss}}(k) = \theta^*(k)$ exponentially as $k\to\infty$ when the initial error $e(0)\neq 0$.}

We will now prove by contradiction that the steady-state trajectory $\theta_{\text{ss}}$ is unique when the agent is not moving on a straight trajectory. 
Let $\theta^{'}_{\text{ss}}\neq \theta_{\text{ss}}$ be another steady-state trajectory satisfying \eqref{eq:steady-state-theta}. 
Since the error is zero for both trajectories, this implies also that 
\begin{equation}
    \label{eq:steady-state}
    %l^* = y(k)\sbm{\cos(\theta_{\text{ss}}(k))\\ \sin(\theta_{\text{ss}}(k))}+p(k) = %y(k)\sbm{\cos(\theta^{'}_{\text{ss}}(k))\\ \sin(\theta^{'}_{\text{ss}}(k))} + p(k),
    l^* = y(k)\bbm{\cos(\theta_{\text{ss}}(k))\\ \sin(\theta_{\text{ss}}(k))}+p(k) = y(k)\bbm{\cos(\theta^{'}_{\text{ss}}(k))\\ \sin(\theta^{'}_{\text{ss}}(k))} + p(k),
\end{equation}
for all $k\geq 0$. 

Let us take a particular $k> 0$. 
By definition, 
\begin{align*}
%    \begin{array}{rl}
    \|l^* - p(k+1)\|^2 =& \|l^* - p(k) - Tu(k)\|^2\\
     =& \|l^{*} - p(k)\|^2 + T^2\| u(k)\|^2 \\
     &%\hspace{0.8cm} 
     - 2T \langle l^{*} - p(k), u(k)\rangle \enspace.
%    \end{array}
\end{align*}
Since the distance to the static landmark is the same for both steady-state trajectories $\theta_{\text{ss}}$ and $\theta_{\text{ss}}^{'}$, substituting both relations in \eqref{eq:steady-state} to the above equation, we can conclude that:
\begin{align*}
    \left\langle  y(k)\sbm{\cos(\theta_{\text{ss}}(k))\\ \sin(\theta_{\text{ss}}(k))}, u(k)\right\rangle 
    &=\left\langle  y(k)\sbm{\cos(\theta^{'}_{\text{ss}}(k))\\ \sin(\theta^{'}_{\text{ss}}(k))}, u(k)\right\rangle \\
    &=\left\langle  l^*-p(k), u(k)\right\rangle.
\end{align*}
%Using the fact that 
Since $l^*=y(k)\sbm{\cos(\theta^*(k))\\ \sin(\theta^*(k))}+p(k)$ it follows that: 
\begin{align*}
    & \left\langle \sbm{\cos(\theta_{\text{ss}}(k))\\ \sin(\theta_{\text{ss}}(k))} - \sbm{\cos(\theta^{'}_{\text{ss}}(k))\\ \sin(\theta^{'}_{\text{ss}}(k))}, u(k)\right\rangle = 0 
\end{align*}
holds for the given $k$. 
If $\theta_{\text{ss}} \neq \theta^{'}_{\text{ss}}$, then the above equations are satisfied only when $u(k)$ is orthogonal to the vector $\sbm{\cos(\theta_{\text{ss}}(k))\\ \sin(\theta_{\text{ss}}(k))} - \sbm{\cos(\theta^{'}_{\text{ss}}(k))\\ \sin(\theta^{'}_{\text{ss}}(k))}\neq 0$.
In the 2D plane, therefore %this means that 
the estimated static landmark 
$y(k)\sbm{\cos(\theta_{\text{ss}}(k))\\ \sin(\theta_{\text{ss}}(k))}+p(k)$ 
mirrors another possible %static 
landmark position 
$y(k)\sbm{\cos(\theta^{'}_{\text{ss}}(k))\\ \sin(\theta^{'}_{\text{ss}}(k))}+p(k)$ 
with respect to the axis collinear with $u(k)$. 

%Let us now consider future time step $k+N$ for some $N> 0$ where $u(k+N)$ is not co-linear with $u(k)$ (by the hypothesis of  proposition that the agent is not moving in a straight line). 
Let us now consider a future time step $k+N$ for some $N>0$, where $u(k+N)$ is not collinear with $u(k)$, i.e., it is not moving on a straight line.
Following the same argument as before we have 
\begin{align*}
    & \left\langle \sbm{\cos(\theta_{\text{ss}}(k+N))\\ \sin(\theta_{\text{ss}}(k+N)} - \sbm{\cos(\theta^{'}_{\text{ss}}(k+N))\\ \sin(\theta^{'}_{\text{ss}}(k+N))},  
    u(k+N)\right\rangle = 0 \enspace. 
\end{align*}
%This leads to a contradiction, since it implies that the alternative static landmark position associated to $\theta^{'}_{\text{ss}}$ has changed its position due to the change of mirror axis that is co-linear with $u(k+N)$. 
This leads to a contradiction, since it implies that the alternative static landmark position associated to $\theta^{'}_{\text{ss}}$ has changed its position, due to the change of the mirror axis that is collinear with $u(k+N)$.
Thus, $\theta_{\text{ss}}$ is unique and corresponds to $l^*$, since $l(k)$ converges to $l^*$ and $e(k)\to 0$ with $k\to \infty$.
\end{proof}
%
%\vspace{0.2cm}
\begin{proposition}
    \label{prop:convergence_straight}
    Assume that the hypotheses in \autoref{prop:contraction} hold for some $c,\gamma > 0$ so that \eqref{eq:estimator_prop} is contracting. 
    If the agent is moving on a straight trajectory, then the estimated landmark position $l(k)$ converges either to the true landmark position $l^*$ or another point $l^{**}$, that is mirrored to $l^*$ with respect to the axis collinear with $u(k)$, as $k\to \infty$.
    \end{proposition}
%\vspace{0.2cm}

\begin{proof}
The proof of the proposition follows the same reasoning as in the proof of \autoref{prop:convergence_no-straight}, where we conclude that there are two distinct points $l^*$ and $l^{**}$ that satisfy \eqref{eq:steady-state} for a given $k$. 
For all future time step $k+N$ with $N>0$ the conclusion remains the same, as $u(k+N)$ is always collinear with $u(k)$.
\end{proof}

\begin{figure}
    \centering
    \def\mySep{0.5cm}
    \begin{tikzpicture}[node distance = 0.1cm,%nodes = {inner sep=0pt},
    every node/.style={minimum size=3pt,inner sep=1pt,font=\footnotesize}, % ,fill=black
    every edge quotes/.style = {auto, font=\footnotesize, sloped}] % ,inner sep=0pt,outer sep=3pt
        \draw (0,0) node[draw,circle,fill=black,anchor=west,label={[below,yshift=-0.1cm]$p(k-1)$}] (pk) {}; % anchor=north east
        \draw (2,0) node[draw,circle,fill=black,anchor=east,label={[below,xshift=-0.4cm,yshift=-0.1cm]$p(k)$}] (pk2) {}; % anchor=north east
        \draw[-stealth] (pk) edge[rectangle,"$Tu(k)$",fill=white] (pk2);
        \draw node[right=of pk2,xshift=2.5cm] (between) {};
        % \draw (5, 1) node[draw,circle,fill=black,anchor=east,label=right:{$l^*$}] (l) {};
        \draw node[draw,circle,fill=black,anchor=east,label=right:{$l^*$},
            above=of between,yshift=\mySep] (l) {};
        \draw (5,-1) node[draw,circle,fill=black,anchor=east,label=right:{$l^{**}$},
            below=of between,yshift=-\mySep] (ls) {};
        \draw[-stealth]  (pk2)  edge[rectangle,"$l^{*}-p(k)$",fill=white] (l);%\draw[->] (pk2) -- (l) node [midway, fill=white] {$l^{**}-p(k+1)$};
        \draw[-stealth]  (pk2)  edge node[draw=none,rectangle,below,fill=white,sloped] {$l^{**}-p(k)$} (ls);% \draw[->] (pk2) -- (ls);
        \draw[dash pattern=on 4pt off 2pt,gray] (l) -- ++(ls);
        \draw[dash pattern=on 4pt off 2pt,gray] (pk2) -- (between);
        % angle
        \pic [draw, angle radius=1cm, angle eccentricity=1,anchor=south west,"$\alpha(k)$"] {angle = ls--pk2--l};
        \node[rectangle,draw,gray,anchor=south east,yshift=-0.07cm] at (between.north) {$\cdot$};
        \node[rectangle,draw,gray,anchor=north east,yshift= 0.07cm] at (between.south) {$\cdot$};
    \end{tikzpicture}
    \caption{Geometric derivation of $\alpha(k)$, the value to which the estimation error $e(k)$ converges when the estimated landmark position $l(k)$ converges to the mirror landmark $l^{**}$.}
    %Angle $\alpha(k)$ denotes the estimation error in case of straight movement $u(k)$ and convergence to the mirror landmark $l^{**}$. %Angle $\alpha(k)$ denotes the estimation error $e(k)$ in case of straight movement $u(k)$ and convergence to the mirror landmark $l^{**}$.
    \label{fig:sketch}
\end{figure}
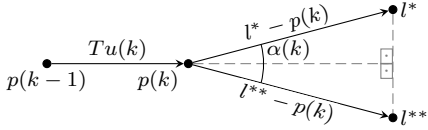
%One can deduce using geometry that when $l(k)$ converges to the other steady-state point $l^{**}$, the estimation error
%\kb{If $l(k)$ converges to the other steady-state point $l^{**}$ we can deduce the estimation error using geometry (see \autoref{fig:sketch}):}
Moreover, using geometry (see \autoref{fig:sketch}) we can conclude that when $l(k)$ converges to the other steady-state point $l^{**}$ then the estimation error $e(k)$ converges to $\alpha(k)$ defined as
\begin{equation*}
    %e(k) \to 
    \alpha(k) = 2\cos^{-1}\left(\dfrac{\langle l^* - p(k),u(k) \rangle}{\|l^* - p(k)\|\|u(k)\|}\right) \enspace.
\end{equation*}

We remark that the above results can further be extended to the case of $m$ different landmarks $l^* = \bbm{l^*_1 & \dots & l^*_m}^{\top}\in \rline^{2m}$. 
In this case, we can extend the proposed estimator \eqref{eq:estimator} straightforwardly where the estimator state is defined by $\theta(k) = \bbm{\theta_1(k) & \dots & \theta_m(k)}^{\top}\in \rline^m$ and the corresponding estimator gain vector $\gamma = \bbm{\gamma_1 & \dots & \gamma_m}^{\top}\in \rline^m$. 
The previous \autoref{prop:contraction}, \ref{prop:convergence_no-straight} and \ref{prop:convergence_straight} still hold separately for each landmark $l^*_i$ with $i = 1,\dots,m$. 
Note that, although it can be challenging to choose a common gain $\gamma>0$ so that the inequalities \eqref{eq:gain_inequality} are satisfied for all the landmarks $l^*_i$, it is possible to use the knowledge on individual distance measurements $y_i(k) = \|l^*_i - p(k)\|$ to select the magnitude of each $\gamma_i$ for all $i = 1,\dots,m$. 

\section{Numerical Simulation}
\label{sec:numerical_sim}
\begin{figure}[t!]
    \centering
    \includegraphics[width=\linewidth]{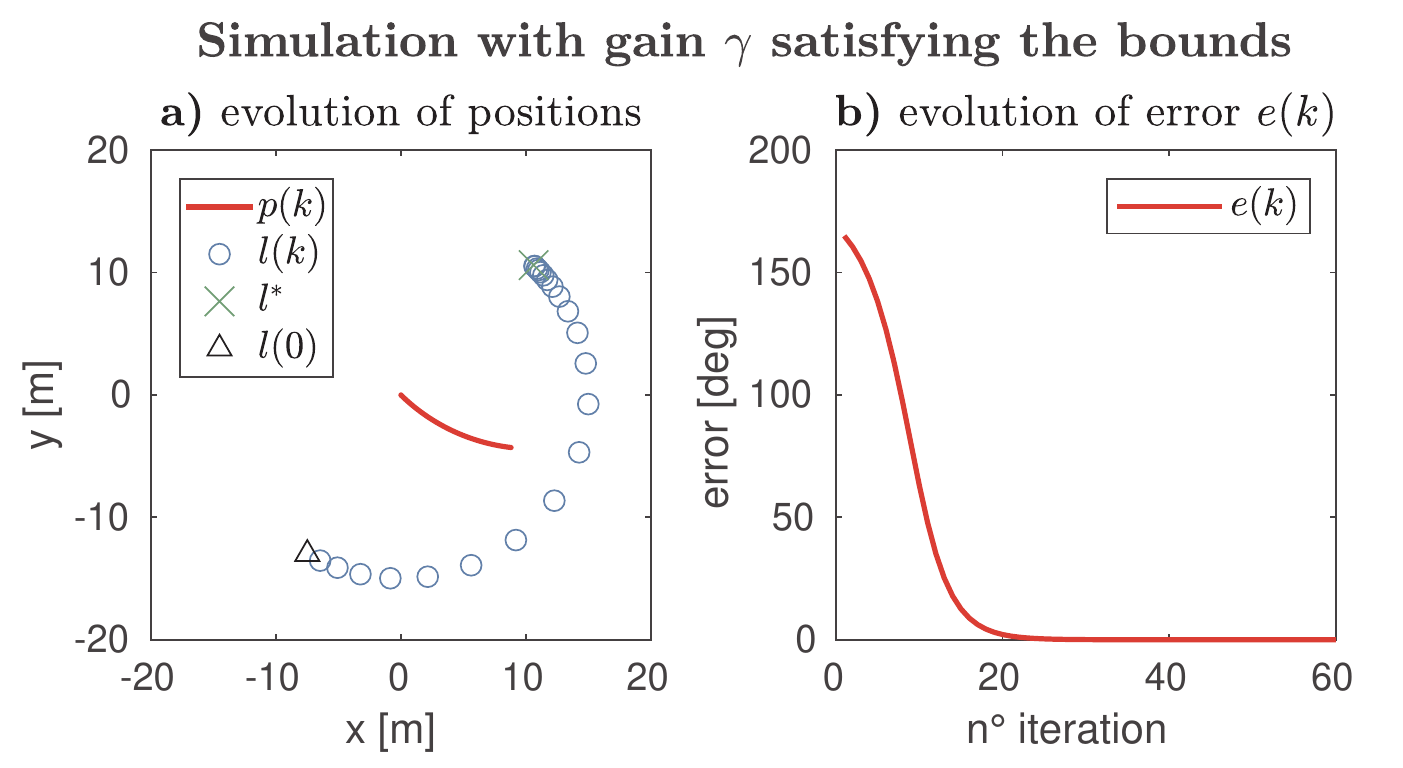}
    %\begin{subfigure}{.5\columnwidth}
    %  \centering
    %  \includegraphics[width=1\columnwidth]{images/position_in-bounds.pdf}
    %  \caption{Evolution of positions over time.} %$p(k),\hatl(k)$ and $l^*$
    %  \label{subfig:position_in-bounds}
    %\end{subfigure}%
    %\begin{subfigure}{.5\columnwidth}
    %  \centering
    %\includegraphics[width=1\columnwidth]{images/error_in-bounds.pdf}
    %\caption{Estimation error over time.}
    %  \label{subfig:error_in-bounds}
    %\end{subfigure}
    \caption{%\printlen[5][cm]{\columnwidth}
        Simulation of curved trajectory for position estimation of a static landmark when the gain $\gamma$ satisfies the bounds: (a) depicts the agent trajectory $p(k)$ (red line), the true location of the landmark $l^*$ (green cross), the estimated landmark position $l(k)$ over time (blue circles), and its initialization $l(0)$ (black triangle); (b) displays the error trajectory $e(k)$.} %the red line represents the agent trajectory $p(k)$, the green cross the true location of the landmark $l^*$ and the blue circles the estimated landmark position trajectory $l(k)$;
    \label{fig:simulation_in-bounds}
\end{figure}

\begin{figure}[t!]
    \centering
    \includegraphics[width=\linewidth]{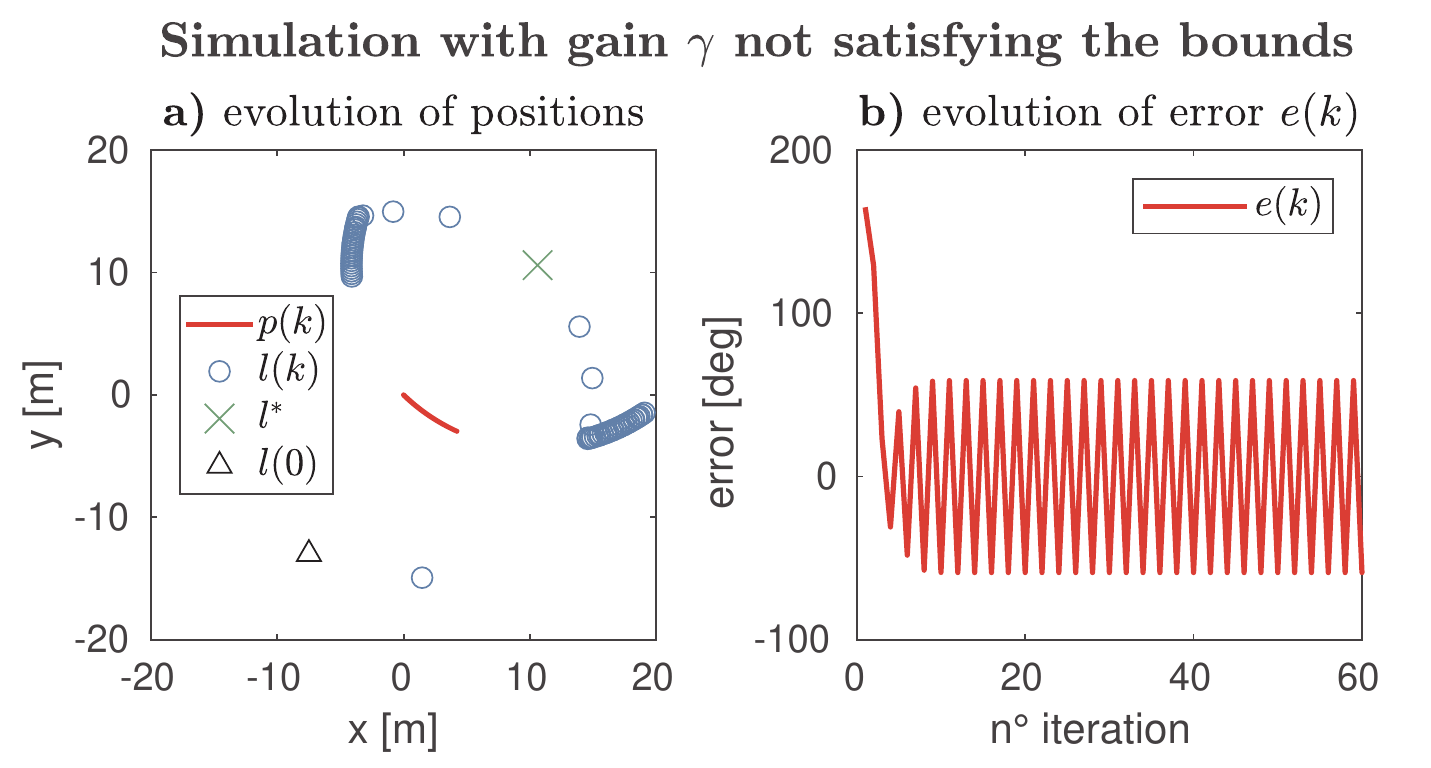}
    %\begin{subfigure}{.5\columnwidth}
    %  \centering
    %  \includegraphics[width=1\columnwidth]{images/position_out-bounds.pdf}
    %  \caption{Evolution of positions over time.}
    %  \label{subfig:position_out-bounds}
    %\end{subfigure}%
    %\begin{subfigure}{.5\columnwidth}
    %  \centering
    %\includegraphics[width=1\columnwidth]{images/error_out-bounds.pdf}
    %\caption{Estimation error over time.}
    %  \label{subfig:error_out-bounds}
    %\end{subfigure}
    \caption{Simulation of curved trajectory for position estimation of a static landmark when the gain $\gamma$ does not satisfies the bounds: (a) depicts the agent trajectory $p(k)$ (red line), the true location of the landmark $l^*$ (green cross), the estimated landmark position $l(k)$ over time (blue circles), and its initialization $l(0)$ (black triangle); (b) displays the error trajectory $e(k)$.}
    %Simulation of \kb{curved} %non straight 
    %trajectory for position estimation of a static landmark when the gain $\gamma$ does not satisfy the bounds: 
    %(a) \kb{depicts the agent trajectory $p(k)$ (red), true location of the landmark $l^*$ (green cross) and the estimated landmark position $l$ over time $k$ (blue circles).}
    %the red line represents the agent trajectory $p(k)$, the green cross the true location of the landmark $l^*$ and the blue circles the estimated landmark position trajectory $l(k)$; 
    %\kb{Panel} (b) %the red line 
    %displays the error trajectory $e(k)$.
    \label{fig:simulation_out-bounds}
\end{figure}

In this section, we present the results of a set of numerical simulations implemented in MATLAB. 
In the first two simulations we evaluated the bounds \eqref{eq:gain_inequality} of the estimator gain $\gamma$ \eqref{eq:estimator} as presented in \autoref{prop:contraction} for a curved trajectory. 
This simulation result is shown in \autoref{fig:simulation_in-bounds}, in which we chose a gain $\gamma = 10$ that satisfies \eqref{eq:gain_inequality} for all $k\geq0$. 
As expected from \autoref{prop:contraction} and \ref{prop:convergence_no-straight} the estimated landmark position converges to the true landmark, i.e. $l(k)\to l^*$ as $k\to \infty$.  
\autoref{fig:simulation_out-bounds} shows instead the simulation when $\gamma = 80$ not satisfying \eqref{eq:gain_inequality}. 
As predicted from the theoretical analysis the estimated landmark position $l(k)$ does not converge to the true landmark $l^*$. 

\begin{figure}[t!]
    \centering
    \includegraphics[width=\linewidth]{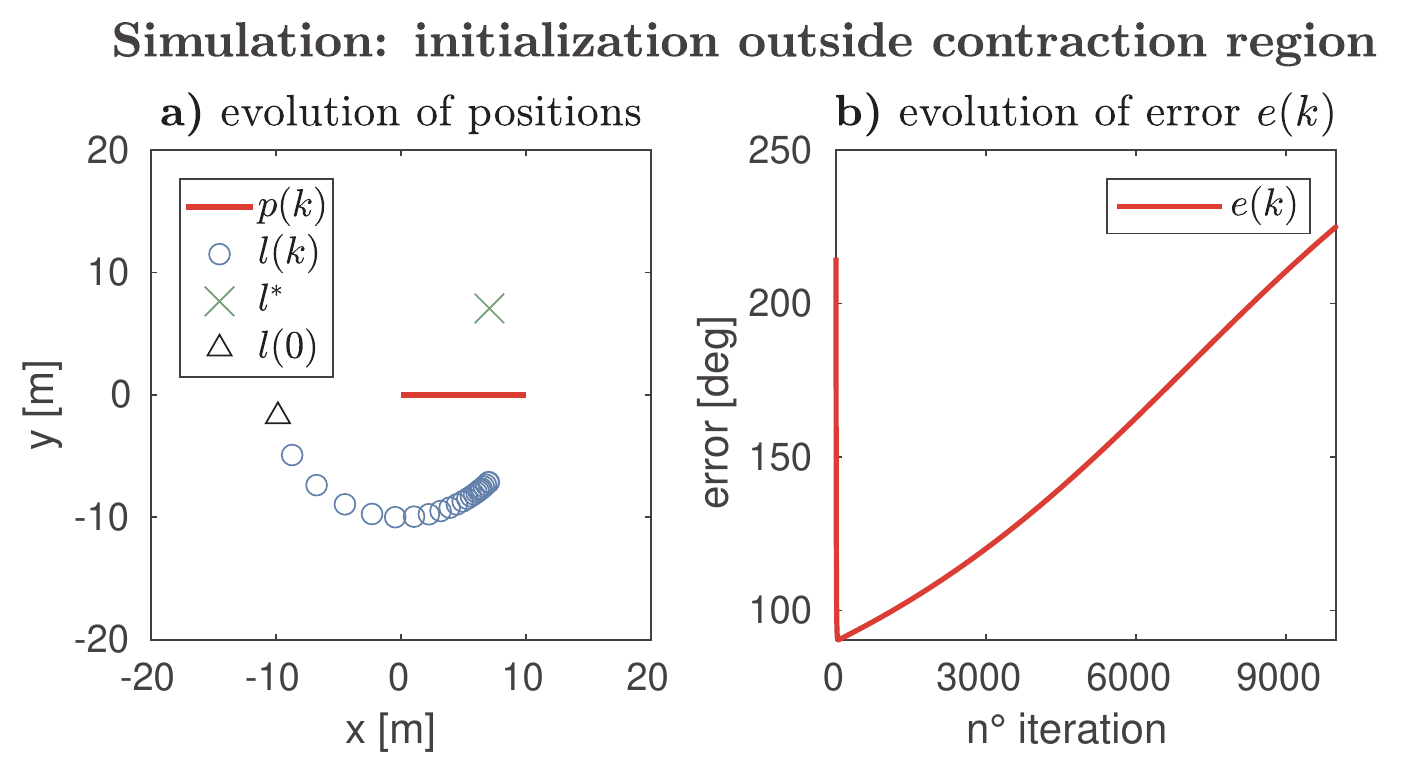}
    %\begin{subfigure}{.5\columnwidth}
    %  \centering
    %  \includegraphics[width=1\columnwidth]{position_straight-out}
    %  \caption{Evolution of positions over time.}
    %  \label{subfig:position_straight-out}
    %\end{subfigure}%
    %\begin{subfigure}{.5\columnwidth}
    %  \centering
    %\includegraphics[width=1\columnwidth]{error_straight-out}
    %\caption{Estimation error over time.}
    %  \label{subfig:error_straight-out}
    %\end{subfigure}
    \caption{Simulation of a straight trajectory when the estimated bearing initialization $\theta(0)$ is outside the contraction region of $l^{*}$: (a) shows the agent trajectory $p(k)$ (red line), the true location of the landmark $l^*$ (green cross), the estimated landmark position trajectory $l(k)$ (blue circles), and and its initialization $l(0)$ (black triangle); (b) displays the error trajectory $e(k)$ with time.}
    \label{fig:simulation_straight-out}
\end{figure}

In the simulation shown in \autoref{fig:simulation_straight-out} we validate the results of \autoref{prop:convergence_straight} on the presence of another attracting point for collinear trajectories $p(k)$. 
It displays the simulation results when $\theta(0)= -170\degree$, such that $\theta(0)$ is outside the contraction region of $l^*$.
As predicted, the estimated landmark position $l(k)$ does not converge to the true landmark $l^*$ but to to the mirror position $l^{**}$. 
%\begin{figure}[t!]
    %\centering
    %\begin{subfigure}{.5\columnwidth}
    %  \centering
      %\includegraphics[width=1\columnwidth]{images/position_straight-in.pdf}
      %\caption{Evolution of positions over time.}
      %\label{subfig:position_straight-in}
    %\end{subfigure}%
    %\begin{subfigure}{.5\columnwidth}
    %  \centering
    %\includegraphics[width=1\columnwidth]{images/error_straight-in.pdf}
    %\caption{Estimation error over time.}
    %  \label{subfig:error_straight-in}
    %\end{subfigure}
    %\caption{Simulation of straight trajectory for position estimation of a static landmark when the estimated bearing initialization $\theta(0)$ is on the contraction region of $l^{*}$: (a) the red line represents the agent trajectory $p(k)$, the green cross the true location of the landmark $l^*$ and the blue circles the estimated landmark position trajectory $l(k)$; (b) the red line displays the error trajectory $e(k)$.}
    %\label{fig:simulation_straight-in}
%\end{figure}

\begin{figure}[t!]
    \centering
    \includegraphics[width=\linewidth]{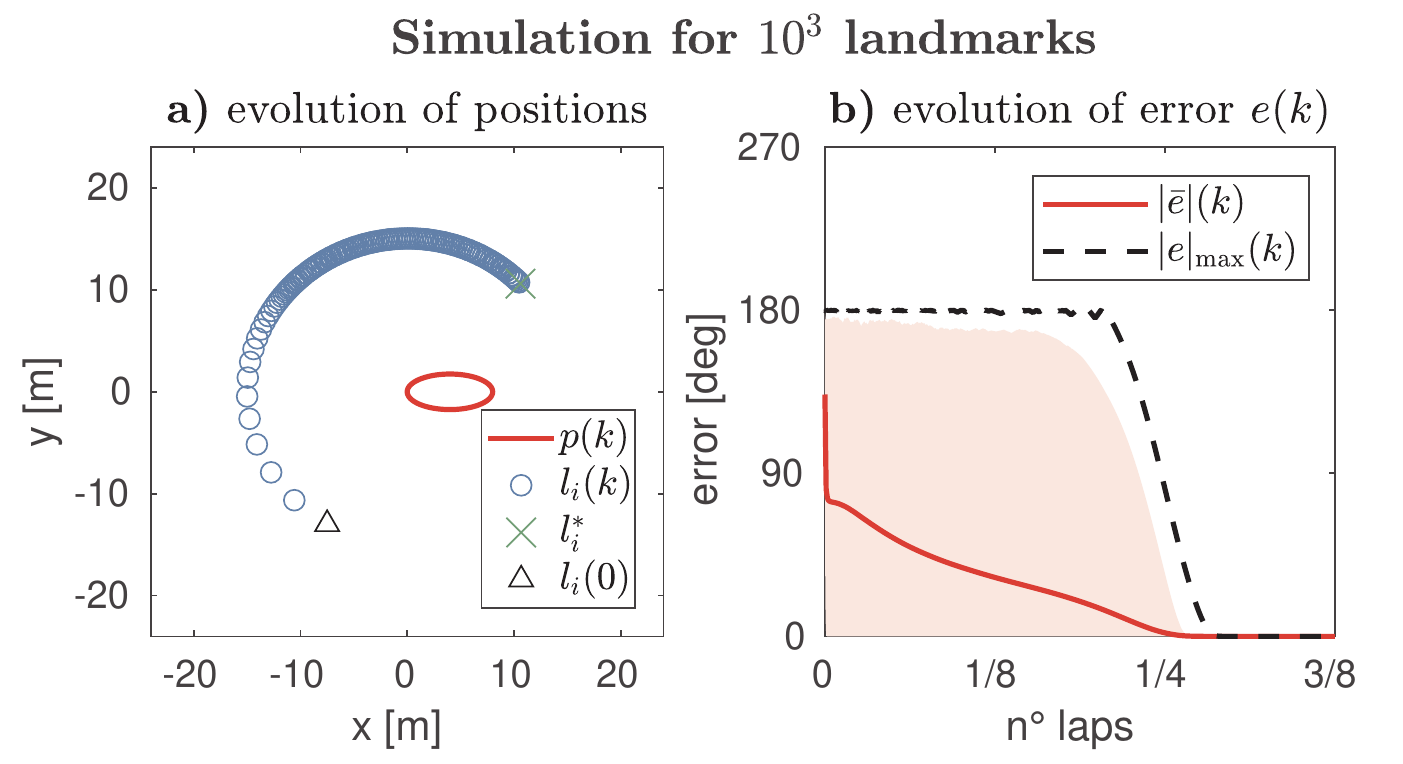}
    %\begin{subfigure}{.5\columnwidth}
    %  \centering
    %  \includegraphics[width=1\columnwidth]{images/multi_p_rounds.pdf}
    %  \caption{Evolution of positions over time.}
    %  \label{subfig:position_multiple-land}
    %\end{subfigure}%
    %\begin{subfigure}{.5\columnwidth}
    %  \centering
    %\includegraphics[width=1\columnwidth]{images/multi_e_75_rounds.pdf}
    %\caption{Estimation error over time.}
    %  \label{subfig:error_multiple-land}
    %\end{subfigure}
    \caption{Simulation for $10^3$ landmarks in case of a closed ellipse trajectory: (a) shows the agent trajectory $p(k)$ (red line), the true location of one landmark $l^*_i$ (green cross), the correspondent estimated landmark position trajectory $l_i(k)$ (blue circles), and its initialization $l_i(0)$ (black triangle); (b) depicts the average $|\bar{e}|(k)$ (red line) and the maximum $|e|_{\max}(k)$ (black dashed line) absolute values of the error over the number of laps, together with its $95\%$ confidence interval plot.}% with red line displaying average absolute value of error over the number of laps.
    \label{fig:simulation_multiple-land}
\end{figure}

Finally in \autoref{fig:simulation_multiple-land} we present results from simulations in case of multiple landmarks. 
While the results from \autoref{prop:contraction} and \ref{prop:convergence_no-straight} are applicable only to the motion in the cones with axis $w_i(k)$ defined for each landmark $i$, we will evaluate the effectiveness of the proposed estimator when the mobile agent completes a closed trajectory. 
Particularly, the agent follows a closed ellipse trajectory for multiple rounds.
%In this example the number of landmarks is $10^3$ and their position has been randomly generated such that their distance from the origin $y_i(0)$ is within the set $[10,25]$ and their bearing from the origin $\theta^*_i(0)$ is within the set $[-\pi,\pi]$ for all $i =1 ,\dots,m$; the same is valid for the initialization of the estimated relative bearings $\theta_i(0)$. 
In this example the number of landmarks is $10^3$ and their positions were randomly generated inside a ring (with inner radius $10$ and outer radius $25$), which is centred with the closed ellipse trajectory.
The estimated relative bearings have also been initialized randomly so that $\theta_i(0)\in [-\pi,\pi]$ for all $i =1 ,\dots,m$.
As estimator gains we selected $\gamma_i = 5,\ \forall i = 1, \dots, m$, such that the contraction condition \eqref{eq:gain_inequality} is satisfied for each landmark.

In \autoref{fig:simulation_multiple-land}a) the evolution over time for one estimated landmark position $l_i(k)$ is shown to converge to the correspondent true landmark $l^*_i$. 
Moreover, \autoref{fig:simulation_multiple-land}b) displays the $95\%$ confidence plot for the absolute values of the estimated errors $\left|e_i(k) \right|,\ \forall i=1,\dots,m$ over the number of laps of the agent following the closed ellipse trajectory, $|\bar{e}|(k)$ represents the average and $|e|_{\max}(k)$ the maximum error trajectories.
Note that all errors converge to zero before even half a lap is completed. 

%\vspace{0.3cm}
\newpage
\section{Conclusion}
\label{sec:conclusions}
%\vspace{-0.7cm}
%This contribution proposes a dynamic bearing estimator that is updated using range and local displacement sensor systems.
This contribution proposes a dynamic bearing estimator that relies exclusively on range and local displacement sensor systems.
Under some sufficient conditions on the estimator gain, we have shown the contraction property of the bearing estimator and its convergence to the actual static landmark position or its mirror dependent on the trajectory of the mobile agent. 
The theoretical findings are demonstrated in several empirical simulations with varying trajectories, estimator gains, as well as a multi-landmark scenario.
Future work will address the generalization of the approach to dynamic landmarks and its application for multi-agent distributed control systems.

\bibliographystyle{IEEEtran}
\bibliography{rootBIB}

\end{document}